\documentclass[twocolumn,showpacs,aps,prl,superscriptaddress,floatfix]{revtex4}
\usepackage{graphicx}
\usepackage{verbatim}
\usepackage{dcolumn}
\usepackage{amsmath}
\usepackage{epsfig}
\usepackage{subfigure}

\newcommand{\BaBarYear}      {16}
\newcommand{\BaBarNumber}    {001}
\newcommand{\BaBarType}      {PUB}  
\newcommand{\SLACPubNumber}  {16526}

\input babarsym.tex

\def\epem{e^+e^-}

\def\Btag{\ensuremath{B_{\rm tag}}\xspace}  

\def \Bsig{\ensuremath{B_{\rm sig}}\xspace}

\begin{document}

\pagestyle{plain}

\begin{flushleft}
\babar-\BaBarType-\BaBarYear/\BaBarNumber \\
SLAC-PUB-\SLACPubNumber\\

\end{flushleft}

\title{{\large \bf Search for \Bp\to\Kp\tautau at the \babar{} experiment}}

% NOTES
%
% 20-FEB-2016 Add footnote for Liang Sun                               J.W. Gary
% 21-DEC-2015 Add Bologna alternative address for Claudia Patrignani   J.W. Gary
%
\author{J.~P.~Lees}
\author{V.~Poireau}
\author{V.~Tisserand}
\affiliation{Laboratoire d'Annecy-le-Vieux de Physique des Particules (LAPP), Universit\'e de Savoie, CNRS/IN2P3,  F-74941 Annecy-Le-Vieux, France}
\author{E.~Grauges}
\affiliation{Universitat de Barcelona, Facultat de Fisica, Departament ECM, E-08028 Barcelona, Spain }
\author{A.~Palano}
\affiliation{INFN Sezione di Bari and Dipartimento di Fisica, Universit\`a di Bari, I-70126 Bari, Italy }
\author{G.~Eigen}
%\author{B.~Stugu}
\affiliation{University of Bergen, Institute of Physics, N-5007 Bergen, Norway }
\author{D.~N.~Brown}
%\author{L.~T.~Kerth}
\author{Yu.~G.~Kolomensky}
%\author{M.~J.~Lee}
%\author{G.~Lynch}
\affiliation{Lawrence Berkeley National Laboratory and University of California, Berkeley, California 94720, USA }
\author{H.~Koch}
\author{T.~Schroeder}
\affiliation{Ruhr Universit\"at Bochum, Institut f\"ur Experimentalphysik 1, D-44780 Bochum, Germany }
% ...
\collaboration{The \babar\ Collaboration}
\noaffiliation


\begin{abstract}

We search for the rare flavor-changing neutral current process \Bp\to$K^{+}$\tautau using data from the \babar{} experiment. The data sample, collected at the center-of-mass energy of the \FourS resonance, corresponds to a total integrated luminosity of 424\invfb and to 471 million \BB pairs. We reconstruct one \B meson, produced in the \FourS\to\BpBm decay, in one of many hadronic decay modes and search for activity compatible with a \Bp\to$K^{+}$\tautau decay in the rest of the event. Each $\tau$ lepton is required to decay leptonically into an electron or muon and neutrinos. Comparing the expected number of background events with the data sample after applying the selection criteria, we do not find evidence for a signal. The measured branching fraction is ($1.31^{+0.66}_{-0.61}$(\rm stat.)$^{+0.35}_{-0.25}$(\rm sys.)$)  \times 10^{-3}$  with an upper limit, at the 90\% confidence level, of \mbox{$\BR(\Bp\to K^{+}\tautau)$}$< 2.25\times 10^{-3}$. 

\end{abstract}

\maketitle

\setcounter{footnote}{0}

\begin{figure}
\begin{center}
\includegraphics[height=4cm, width=7.5cm]{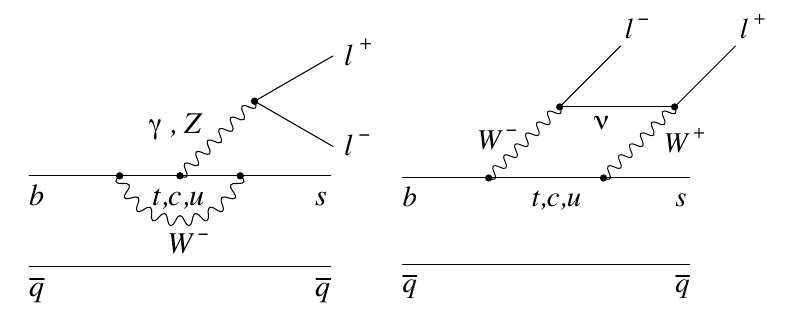}
\caption{Lowest order SM Feynman diagrams of \b\to\s\ellell.}
\label{fig1}
\end{center}
\end{figure}

The flavor-changing neutral current process \mbox{\Bp\to\Kp\tautau}~\cite{cc} is highly suppressed in the standard model (SM), with a predicted branching fraction in the range $1-2 \times 10^{-7}$~\cite{lattice,hewitt}.  This decay is forbidden at tree level and only occurs, at lowest order, via one-loop diagrams. The SM contributions, shown in Fig.~\ref{fig1}, include the electromagnetic penguin, the $Z$ penguin, and the $W^{+}W^{-}$ box diagrams. Rare semi-leptonic \B decays such as \mbox{\Bp\to$K^{+}$\tautau} can provide a stringent test of the SM and a fertile ground for new physics searches. Virtual particles can enter in the loop and thus allow to probe, at relatively low energies, new physics at large mass scales. Measurements of the related decays, \mbox{\Bp\to$K^{+}$\ellell} where $\ell= e$ or $\mu$, have been previously published by \babar{}~\cite{babarBtoKll} and other experiments~\cite{otherBtoKll}, and exhibit some discrepancy with the SM expectation~\cite{review1}.

The decay \Bp\to$K^{+}$\tautau is the third family equivalent of \Bp\to$K^{+}$\ellell and hence may provide additional sensitivity to new physics due to third-generation couplings and the large mass of the $\tau$ lepton~\cite{review}. An important potential contribution to this decay is from neutral Higgs boson couplings, where the lepton-lepton-Higgs vertices are proportional to the mass squared of the lepton~\cite{aliev}. Thus, in the case of the $\tau$, such contributions can be significant and could alter the total decay rate. Additional sources of new physics and their effect on the \mbox{\Bp\to$K^{+}$\tautau} branching fraction and the kinematic distributions of the \tautau pair are also discussed in Refs.~\cite{su5}-\cite{newRef}.

We report herein a search for \Bp\to$K^{+}$\tautau with data recorded by the \babar\ detector~\cite{Bib:Babar} at the  $ \epem$ \pep2 collider at the SLAC National Accelerator Laboratory. This search is based on 424 fb$^{-1}$ of data~\cite{lum} collected at the center-of-mass (CM) energy of the $\FourS$ resonance, where \FourS decays into a \BB pair. We use hadronic \B meson tagging techniques, where one of the two \B mesons, referred to as the \Btag, is reconstructed exclusively via its decay into one of several hadronic decay modes. The remaining tracks, clusters, and missing energy in the event are attributed to the signal \B, denoted as \Bsig, on which the search for \mbox{\Bp\to\Kp\tautau} is performed. We consider only leptonic decays of the $\tau$: $\taup\to\ep\nue\nutb$ and $\taup\to\mup\num\nutb$, which results in three signal decay topologies with a charged \kaon, multiple missing neutrinos, and either $\epem$, $\mumu$ or $\ep\mun$ in the final state. The neutrinos are accounted for as missing energy in any signal event where a charged kaon and lepton pair are identified and extra neutral activity, including \piz candidates, is excluded.

Simulated Monte Carlo (MC) signal and background events, generated with EvtGen~\cite{evtgen}, are used to develop signal selection criteria and to study potential backgrounds. The  detector response is simulated using GEANT4~\cite{Geant4}. Signal MC events are generated as \mbox{\FourS\to\BpBm}, where one \B decays according to its measured SM branching fractions~\cite{pdg} and the other \B decays via \mbox{\Bp\to$K^{+}$\tautau} according to the model described in Ref.~\cite{ali}. Within this model, referred to as LCSR, a light-cone sum rule approach is used to determine the form factors  that enter into the parameterization of the matrix elements describing this decay. Signal events are also reweighted to a model based on the unquenched lattice QCD calculations of the \mbox{\B\to$K$\ellell} form factors \cite{lattice} for the determination of the signal efficiency, and the two theoretical approaches are then compared to evaluate the model-dependence of our measurement. Because of the low efficiency of the hadronic \Btag reconstruction, ``dedicated" signal MC samples are also generated for this analysis, where one \B decays exclusively through $\Bpm\to\Dz\pi^{\pm}, \Dz\to K^{-}\pip$ while the other \B meson decays via the signal channel. This ensures that more events pass the hadronic \Btag reconstruction and allows for increased statistics in the distributions of discriminating variables in the signal sample. Only variables that are independent of the \Btag decay mode are considered with the dedicated signal MC sample. To avoid potential bias, this dedicated sample is not used to evaluate the final signal selection efficiency. Background MC samples consist of \BpBm and \BzBzb decays and continuum events, $\epem\to f\bar{f}$, where $f$ is a lepton or a quark. The \BB and $\epem\to\ccbar$ MC-simulated samples are produced with an integrated luminosity ten times that of data, whereas the remaining continuum samples have an integrated luminosity that is four times larger. 

The signal selection of \Bp\to$K^{+}$\tautau events is preceded by the full hadronic reconstruction of the \Btag meson, via  \B\to$S X$~\cite{knunuPaper}. Here, $S$ is a seed meson, $D^{(*)0}$, $D^{(*)\pm}$,  $D^{*\pm}_{s}$ or \jpsi, and $X$ is a combination of at most five kaons and pions with a total charge of 0 or $\pm1$. The $D$ seeds are reconstructed in the decay modes $D^+\to\KS\pip$, $\KS\pip\piz$, $\KS\pip\pim\pip$, $ K^-\pip\pip$, $K^-\pip\pip\piz$, $K^{+}\Km\pip$, $K^{+}\Km\pip\piz$; 
$\Dz\to K^-\pip$, $K^-\pip\piz$, $K^-\pip\pim\pip$, $\KS\pip\pim$, $\KS\pip\pim\piz$, $K^{+}\Km$, $\pip\pim$, $\pip\pim\piz$, and $\KS\piz$; 
$D^{*+}\to \Dz\pip$, $\Dp\piz$; $\Dstarz\to \Dz\piz$, $\Dz\g$. The \Dss and \jpsi seeds are reconstructed as $D_s^{*+}\to D_s^+\g$; $D_s^+\to\phi\pip$, $\KS K^{+}$; and $J/\psi\to\epem$, $\mu^+\mu^-$, respectively. \KS and $\phi$ candidates are reconstructed via their decay to $\pip\pim$ and $K^{+}K^{-}$, respectively.

We select \Btag candidates using two kinematic variables: $\mes =\sqrt{(E^{*}_{\rm CM}/2)^2-\vec{p^{*}}_{ \Btag}^2}$ and $\Delta E=\frac{E^{*}_{\rm CM}}{2}-E^{*}_{ \Btag}$, where $E^{*}_{ \Btag}$  and $\vec{p}^{*}_{ \Btag}$ are the CM energy and three-momentum vector of the \Btag, respectively, and $\frac{E^{*}_{\rm CM}}{2}$ is the CM beam energy. A properly reconstructed \Btag has \mes consistent with the mass of a \B meson and $\Delta E$ consistent with 0 \gev. We require $5.20<\mes<5.30 \gevcc$ and $-0.12<\Delta E<0.12 \gev$, where the \mes range includes a sideband region for background studies. If more than one \Btag candidate per event satisfies these requirements, the \Btag candidate in the highest purity mode is chosen.  The purity of a \Btag decay mode is determined from MC studies and is defined as the fraction of \Btag candidates with $\mes>5.27 \gevcc$ that are properly reconstructed within the given mode. If more than one \Btag candidate with the same purity exists,  the one with the smallest $\lvert \DeltaE \lvert$ is chosen.

The hadronic \Btag reconstruction results in both charged and neutral \B mesons. Since the \Btag is fully reconstructed, its four-vector is fully determined and thus that of the \Bsig can be calculated. The latter is obtained using  $|\vec{p^{*}}_{ \Bsig}| = \sqrt{(E^{*}_{ \rm CM}/2)^2 - m^2_{\B}}$, where $\vec{p^{*}}_{ \Bsig}$ is the three momentum vector of \Bsig in the CM frame and $m_{B}$ is the mass of the \B meson, with the direction of  $\vec{p}^{*}_{ \Bsig}$ opposite to that of $\vec{p}^{*}_{ \Btag}$. The missing momentum four-vector, $p^{*}_{\rm miss}$, is determined by subtracting the CM four-momentum of all ``signal-side" tracks and clusters from that of the \Bsig. 

\Bp\to$K^{+}$\tautau signal events are required to have a  charged \Btag candidate with $\mes>5.27 \gevcc$ and a non-zero missing energy, $E_{\rm miss}$, given by the energy component of $p^{*}_{\rm miss}$. Furthermore, to reduce contamination from mis-reconstructed events with high-multiplicity \Btag decay modes, the purity of \Btag candidates is recalculated at this point after also requiring that there remain only three charged tracks in the event not used in the \Btag reconstruction (corresponding to the track multiplicity in signal events). This purity is more relevant to the signal selection, since only charged \Btag decay modes reconstructed with low multiplicity \Bsig events are considered. Signal events with a purity  greater than 40\% are retained. 

Continuum events are further suppressed using a multivariate likelihood selector, which consists of six event-shape variables. These include the magnitude of the \Btag thrust,  defined as the axis which maximizes the sum of the longitudinal momenta of an event's decay products, and  its component along the beam axis and the ratio of the second-to-zeroth Fox-Wolfram moment~\cite{wolfram}. The remaining variables are the angle of the missing momentum vector, $\vec{p^{*}}_{\rm miss}$, with the beam axis,  the angle between $\vec{p}^{*}_{ \Btag}$ and the beam axis, and the angle between the thrust axis of the \Btag and that of the \Bsig in the CM frame. The six event-shape variables discriminate between  \BB events, where the spin-zero \B mesons are produced almost at rest and the decay daughters consequently produce an isotropic distribution,  and continuum events. In the latter, fermions are initially produced with higher momentum, resulting in a more collinear distribution of the final decay products. We require the likelihood ratio\\
\begin{equation}
\mathcal{L}=\frac{\prod\nolimits_i P_{\B} (x_i)}{\prod\nolimits_i P_{\B}(x_i) +\prod\nolimits_i P_{q}(x_i)}>0.50,
\label{eq1}
\end{equation}
where $P (x_i)$ are probability density functions, determined from MC samples, that describe the six event shape variables for \BB, $P_{\B} (x_i)$, and continuum, $P_{\q} (x_i)$, events. This requirement removes more than 75\% of the continuum events while retaining more than 80\% of (signal  and  background) \BB MC events.

A signal selection is then applied on the charged tracks and neutral clusters that are not used in the \Btag reconstruction. \Bp\to$K^{+}$\tautau candidates  are required to possess exactly three charged tracks satisfying particle identification (PID) requirements consistent with one charged \kaon and an $\epem$, $\mumu,$ or $e^{+}\mu^{-}$ pair. The PID selection algorithms for charged tracks are based on multivariate analysis techniques that use information from the \babar{} detector subsystems~\cite{pid}. The \Kpm is required to have a charge opposite to that of \Btag. Furthermore, events with $3.00<m_{\ellell}<3.19 \gevcc$ are discarded to remove backgrounds with a \jpsi resonance. The invariant mass of the combination of the \kaon with the oppositely charged lepton must also lie outside the region of the \Dz mass, i.e. $m_{\Km\ellp}<1.80$\gevcc or $m_{\Km\ellp}>1.90 \gevcc$, to remove events where a pion coming from the \Dz decay is misidentified as a muon. Moreover, events with $\gamma\to\epem$ are  removed by requiring the invariant mass of each electron with any other oppositely charged track in the event to be greater than 50 \mevcc. Background events with \piz candidates,  reconstructed from a pair of photons with individual energies greater than 50 \mev, a total CM energy greater than 100 \mev, and an invariant mass ranging between 100 and 160 \mevcc, are rejected. Additional calorimeter clusters not explicitly associated with \Btag daughter particles may originate from other low-energy particles in background events.  We therefore define $E^{*}_{\rm extra}$ to be the energy sum of all neutral clusters with individual energy greater than 50 \mev that are not used in the \Btag reconstruction.

The normalized squared mass of the \tautau pair is given by $s_{B}=(p_{\Bsig} - p_{\kaon})^2/m^2_B$, where $p_{\Bsig}$ and  $p_{\kaon}$ are the four-momentum vectors of \Bsig and of the kaon, respectively, in the laboratory frame.  The large mass of the $\tau$ leptons in signal events kinematically limits the $s_B$ distribution to large values. A requirement of  $s_{B}>0.45$ is applied.

\begin{figure}
\begin{center}
\includegraphics[height=5cm,width=9cm]{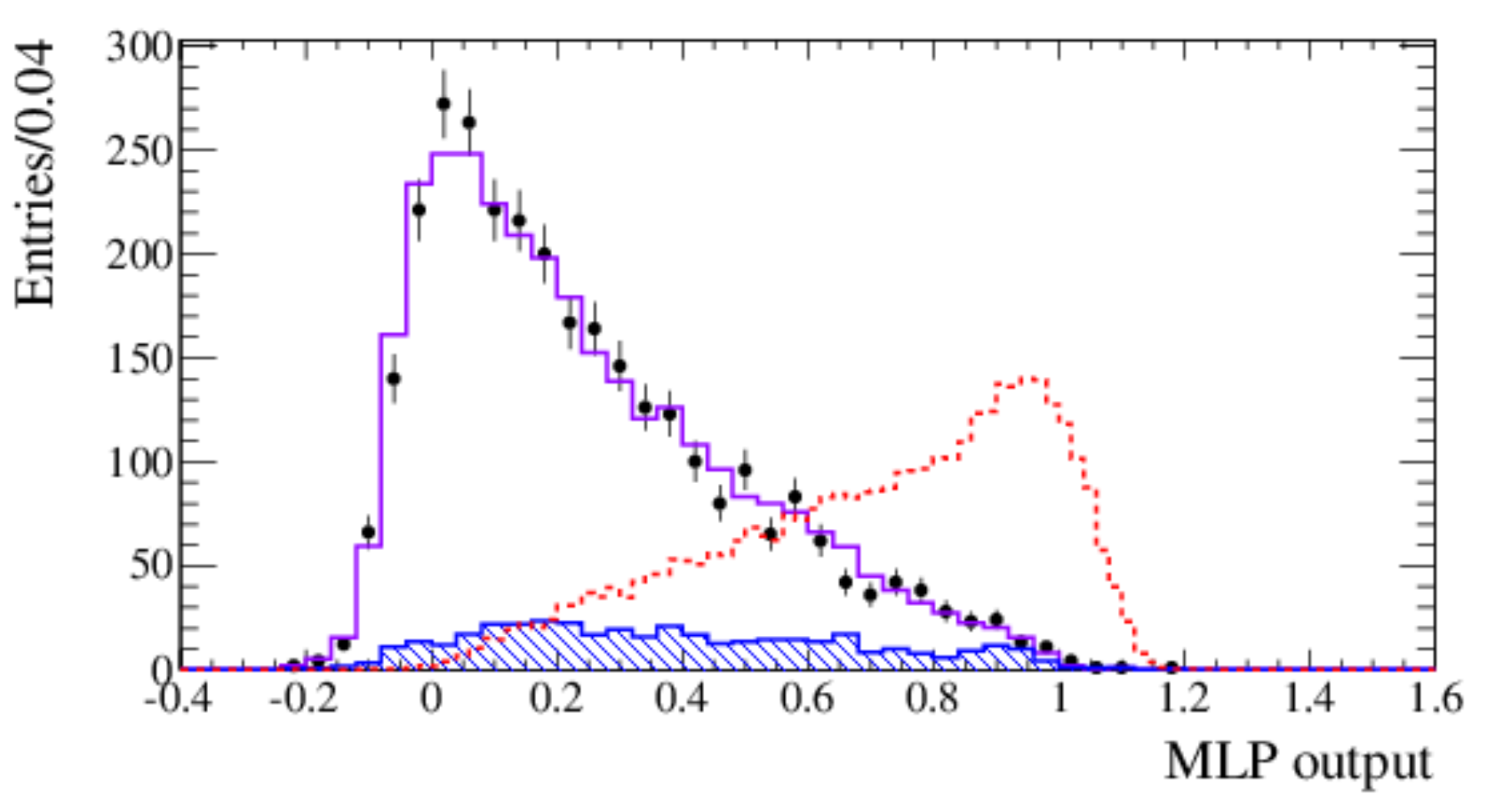}\\
\caption{(color online) MLP output distribution for the three signal channels combined. The \Bp\to$K^{+}$\tautau signal MC distribution is shown (dashed) with arbitrary normalization. The data (points) are overlaid on the expected combinatorial (shaded) plus \mes-peaking (solid) background contributions.}
\label{fig2}
\end{center}
\end{figure}

At this point in the selection, remaining backgrounds are primarily \BB events in which a properly reconstructed \Btag is accompanied by $\Bsig\to D^{(*)} \ell \nulb$, with $D^{(*)}\to\kaon \ell' \overline{\nu}_{\ell'}$ and thus have the same detected final-state particles as signal events. A multi-layer perceptron (MLP) neural network~\cite{neuralnetwork}, with seven input variables and one hidden layer, is employed to suppress this background. The input variables are:  the angle between the kaon and the oppositely charged lepton, the angle between the two leptons, and  the momentum of the lepton with charge opposite to the \kaon, all in the \tautau rest frame, which is calculated as $p_{\rm \Bsig}-p_{\kaon}$; the angle between the \Bsig and the oppositely charged lepton, the angle between the \kaon and the low-momentum lepton,  and the invariant mass of the $K^{+}\ellm$ pair, all in the CM frame. Furthermore, the final input variables to the neural network are  $E^{*}_{\rm extra}$ and the residual energy, $E_{\rm res}$, which here is effectively the missing energy associated with the \tautau pair and is calculated as the energy component of $p^{\tau}_{\rm residual}=p^{\tau}_{\Bsig}-p^{\tau}_{\kaon}-p^{\tau}_{\ellell}$, where $p^{\tau}_{\Bsig}$, $p^{\tau}_{\kaon}$ and $p^{\tau}_{\ellell}$ are the four-momenta vectors in the \tautau rest frame of the \Bsig, \kaon, and lepton pair in the event, respectively. $E_{\rm res}$ has, in general, higher values for signal events than generic \BB and continuum events due to the higher neutrino multiplicity. A neural network is trained and tested using randomly split dedicated signal MC and \BpBm background events, for each of the three channels: $\epem$, $\mumu,$ and $\ep\mun$. The results are shown in Fig.~\ref{fig2} for the three modes combined. The last step in the signal selection is to require that the output of the neural network is $>0.70$ for  the $\epem$ and $\mumu$ channels and $>0.75$ for the $\ep \mun$ channel. This requirement is optimized to yield the most stringent upper limit in the absence of a signal.

The branching fraction for each of the signal modes, $i$,  is calculated as:
\begin{equation}
\BR_i=\frac{N^i_{\rm obs}-N^i_{\rm bkg}}{\epsilon^i_{\rm sig} N_{\BB}},
\label{eq2}
\end{equation}
where $N_{\BB} =471\times 10^6$ is the total number of \BB pairs in the data sample, assuming equal production of \BpBm and \BzBzb pairs in \FourS decays, and $N^i_{\rm obs}$ is the number of data events passing the signal selection.  The signal efficiency, $\epsilon^i_{\rm sig}$, and the background estimate, $N^i_{\rm bkg}$, are determined for each mode from the signal and background MC yields after all selection requirements. 

For each mode, $N_{\rm bkg}$ consists of two components: background events that have a properly reconstructed \Btag and thus produce a distribution in \mes which peaks at the \B mass, and  combinatorial background events composed of continuum and \BB events with mis-reconstructed \Btag candidates which do not produce a peaking structure in the \mes signal region. After the MLP output requirement, peaking background events comprise more than 92\% of the total $N_{\rm bkg}$ for all three modes. To reduce the dependence on MC simulation, the combinatorial background is extrapolated directly from the yield of data events in the \mes ``sideband" region ($5.20<\mes<5.26 \gevcc$), after the full signal selection. Sideband data events are scaled to  the correct normalization of the combinatorial background in the \mes signal region.

The peaking background is determined using \BpBm background MC, while data in the final signal region is kept blinded to avoid experimentalist bias. Because of the large uncertainties on the branching fractions of many of the \Btag decay modes as well as their associated reconstruction effects, there is a discrepancy in the \Btag yield of approximately 10\% between MC and data, independent of the signal selection. A \Btag yield correction is therefore determined by calculating the ratio of data to \BpBm MC events after the $s_B$ requirement. The data sample after this requirement contains a sufficiently large background contribution after the $s_B$ requirement, which consists mainly of \BpBm events ($>96\%$) according to MC simulation, to allow for a data-driven correction without unblinding the final signal-region. This correction factor is determined to be 0.913 $\pm$ 0.020  and is applied to the MC reconstruction efficiency  for both signal and background events.

\begin{figure}
\begin{center}
\includegraphics[height = 5cm ,width=9cm]{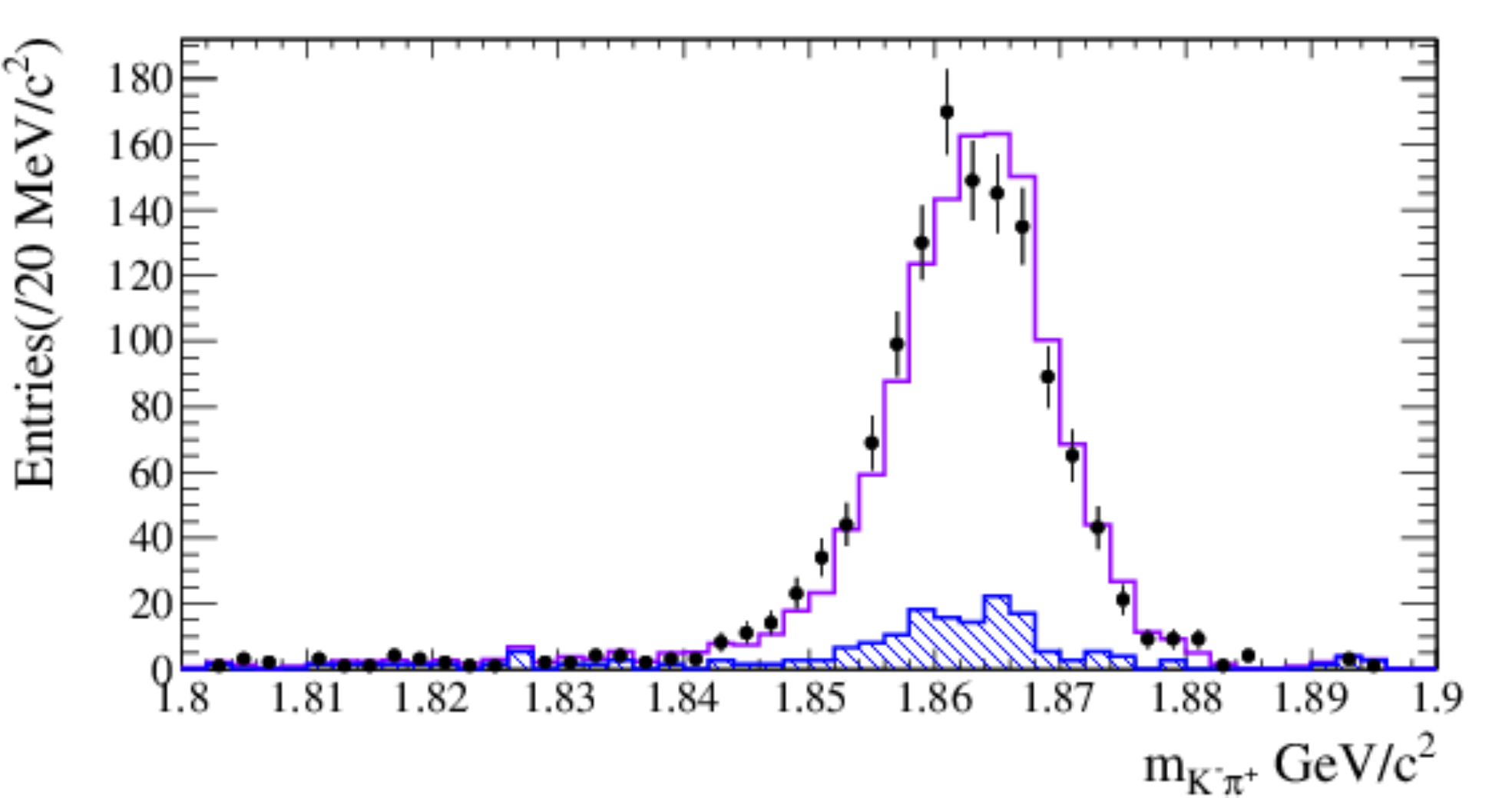}
\caption{(color online) Invariant-mass distribution of the $K^{-}\ell^{+}$ pair in the $\Bp\to\Dz\ellp\nulb, \Dz\to K^{-}\pip$ samples after all signal selection criteria are applied, except for the final requirement on the MLP output. The data (points) are overlaid on the expected combinatorial (shaded) plus \mes-peaking (solid) background contributions.}
\label{fig3}
\end{center}
\end{figure}

The \Btag yield is also cross-checked using a \Bp\to$\Dz\ellp\nul$,  $\Dz\to\Km\pip$ control sample, which is selected using the same signal selection discussed above, but with requiring one track to satisfy pion instead of lepton PID and reversing the \Dz veto, such that $1.80<m_{\Km\pip}<1.90$ \gevcc. These criteria are also applied to the full background MC and the resulting sample is found to consist mainly of peaking \BpBm events, which the MLP neural network is trained to classify as background. Before the MLP requirement, good agreement between data and MC is found in all the distributions of  the input variables of the $\Bp\to\Dz\ellp\nulb, \Dz\to\Km\pip$  samples, as shown in Fig.~\ref{fig3} for the $m_{\Km\pi^{+}}$ distribution. These samples are then run through the MLP neural network and a detailed comparison of the MLP output and the input variables, after the full signal selection, is performed.

The results for each signal channel are then combined to determine \BR(\Bp\to$K^{+}$\tautau). This is done using a frequentist approach by finding the value of  $\BR$ that maximizes the product of the Poisson likelihoods of observing $N^{i}_{\rm obs}$ in each of the signal channels. Branching fraction uncertainties and limits are determined using the method described in Ref.~\cite{barlow}, taking into account the statistical and systematic uncertainties on $N_{\rm bkg}$ and $\epsilon_{\rm sig}$.

Systematic uncertainties associated with the level of data-MC agreement are determined for most of the variables used in the signal selection. The determination of the \Btag yield correction is anti-correlated with the extrapolation of the combinatorial background from the \mes sideband,  as both use the combinatorial background shape from MC. Therefore, only one systematic uncertainty on the \Btag yield and combinatorial background estimate is evaluated, using a simulated MC sample composed of background events with the same luminosity as the data sample. Accounting for the anti-correlation, the effect of varying the value of the \Btag yield correction on the final signal efficiency and background estimate is determined to be $1.2\%$ and $1.6\%$, respectively. The uncertainty associated with the theoretical model is evaluated by reweighting the $s_{B}$ distribution of the dedicated signal MC sample to the LCSR~\cite{ali} theoretical model and to that of Ref.~\cite{dispersion} and determining the difference in signal efficiency, which is calculated to be 3.0\%. Additional uncertainties on $\epsilon_{\rm sig}$ and $N_{\rm bkg}$ arise due to the modeling of PID selectors (4.8\% for $\epem$, 7.0\% for \mumu, and 5.0\% for \ep\mun) and the \piz veto (3.0\%). 
The level of agreement between data and MC is evaluated using the $\Bp\to\Dz\ellp\nul, \Dz\to K^{-}\pip$ control sample before and after the MLP requirement. Comparison of both the overall yields as well as the distributions of the input and output variable results in a systematic uncertainty of 2.6\%. Other potential sources of systematic uncertainties have been investigated, including those associated with the assumption that charged and neutral \B candidates are produced at equal rates, the continuum likelihood suppression, \Btag purity, track multiplicity, $E_{\rm miss}$ and $s_{B}$ selection criteria, and are all implicitly accounted for in the \Btag yield correction uncertainty. Correlations between the signal efficiency and the background estimate due to common systematic errors are included, but are found to have a negligible effect on the final branching fraction results.

\begin{table}
\begin{center}
\begin{tabular}{lcccc}
\hline
\hline
 & $\epem$ & \mumu & \ep\mun \\ \hline \hline

$N^i_{\rm bkg}$   & 49.4$\pm$2.4$\pm$2.9 &45.8$\pm$2.4 $\pm$3.2 &59.2$\pm$2.8 $\pm$3.5 \\
$\epsilon^{i}_{\rm sig} (\times 10^{-5})$& 1.1 $\pm$0.2$\pm$0.1& 1.3$\pm$0.2$\pm$0.1& 2.1$\pm$0.2$\pm$0.2\\
$N^i_{\rm obs}$     & 45&39&92 \\ 
Significance ($\sigma$)&-0.6&-0.9&3.7\\
\hline
\hline

 	\end{tabular}
\caption{Expected background yields, $N^i_{\rm bkg}$, signal efficiencies, $\epsilon^{i}_{\rm sig}$ , number of observed data events, $N^i_{\rm obs}$, and signed significance for each signal mode. Quoted uncertainties are statistical and systematic.}
\label{table1}
	\end{center}
\end{table}

  The final signal efficiencies, background estimates and observed yields of each signal mode are shown in Table~\ref{table1}, with the associated branching fraction significance. The yields in the $\epem$ and \mumu channels show consistency with the expected background estimate. The yields in the \ep\mun channel consist of 40 \ep\mun and 52 \en\mup events, which corresponds to an excess of 3.7$\sigma$ over the background expectation. Examination of kinematic distributions in the \ep\mun channel does not give any clear indication either of signal-like behavior or of systematic problems with background modeling. When combined with the $\epem$ and \mumu modes, the overall significance of the  \Bp\to$K^{+}$\tautau signal is less than $2\sigma$, and hence we do not interpret this as evidence of signal. The  branching fraction for the combined three modes is \mbox{$\BR(\Bp\to K^{+} \tautau)$}=$\large(1.31_{-0.61}^{+0.66}(\rm stat.)_{-0.25}^{+0.35}(\rm sys.)\large) \times 10^{-3}$. The upper limit at the 90\% confidence level is \mbox{$\BR(\Bp\to K^{+}\tautau)$}$<2.25 \times 10^{-3}$.

In conclusion, this is the first search for the decay \mbox{ \Bp\to$K^{+}$\tautau}, using the full \babar{} dataset collected at the CM energy of the \FourS resonance. No significant signal is observed and the upper limit on the final branching fraction is determined to be $2.25  \times 10^{-3}$ at the 90\% confidence level.

\section{Acknowledgments}
\label{sec:Acknowledgments}

We are grateful for the excellent luminosity and machine conditions
provided by our \pep2\ colleagues, 
and for the substantial dedicated effort from
the computing organizations that support \babar.
The collaborating institutions wish to thank 
SLAC for its support and kind hospitality. 
This work is supported by
DOE
and NSF (USA),
NSERC (Canada),
CEA and
CNRS-IN2P3
(France),
BMBF and DFG
(Germany),
INFN (Italy),
FOM (The Netherlands),
NFR (Norway),
MES (Russia),
MINECO (Spain),
STFC (United Kingdom),
BSF (USA-Israel). 
Individuals have received support from the
Marie Curie EIF (European Union)
and the A.~P.~Sloan Foundation (USA).

% NOTES:
% add "and the Binational Science Foundation (U.S.-Israel)"  07-Oct-2013 Bill Gary (Abi Soffer request)


\begin{thebibliography}{9}
\bibitem{cc}
Charge conjugation is implied throughout the entire paper.
\end{thebibliography}
\end{document}